\documentclass[%
 reprint,
superscriptaddress,
showpacs,preprintnumbers,
nobibnotes,
 amsmath,amssymb,
 aps,
prb,
]{revtex4-1}
\usepackage[english]{babel}
\usepackage{graphicx}
\usepackage{dcolumn}
\usepackage{bm}
\usepackage{filecontents}

\usepackage{hyperref}

\begin{document}
\hyphenation{Rehberg}


\title{Dotriacontapoles -- almost self-assembled}

\author{Stefan Hartung}
\email{Stefan.Hartung@uni-bayreuth.de}
\affiliation{Experimentalphysik V, Universit\"at Bayreuth, D-95440 Bayreuth, Germany}
\author{Felix Sommer} 
\affiliation{Experimentalphysik V, Universit\"at Bayreuth, D-95440 Bayreuth, Germany}
\author{Simeon V\"olkel}
\affiliation{Experimentalphysik V, Universit\"at Bayreuth, D-95440 Bayreuth, Germany}

\author{Johannes Sch\"onke}
\email{Johannes.Schoenke@oist.jp}
\affiliation{Okinawa Institute of Science and Technology Graduate University, Onna, Okinawa 904-0495, Japan}
\author{Ingo Rehberg}
\email{Ingo.Rehberg@uni-bayreuth.de}

\affiliation{Experimentalphysik V, Universit\"at Bayreuth, D-95440 Bayreuth, Germany}

\date{ \today}

\begin{abstract}
The magnetic field of a cuboidal cluster of eight magnetic spheres is measured. It decays with the inverse seventh power of the distance. This corresponds formally to a hitherto unheard-of multipole, namely a dotriacontapole. This strong decay is explained on the basis of dipole-dipole interaction and the symmetry of the ensuing ground state of the cuboidal cluster. A method to build such dotriacontapoles is provided.
\end{abstract}
\maketitle

\section{Introduction}
Within the forces determining  the interplay of condensed matter, the dipole-dipole interaction can be considered as the most important one, because monopoles do not exist for neutral matter, and pure quadrupole, octopole or hexadecapol interaction tends to be masked by induced dipole moments. While the interaction of quadrupoles is not too exotic \cite{Buckingham1959} and includes examples from continuum mechanics \cite{Stamou2000},  pure octopole or even higher order interaction has never been reported. Here we demonstrate that the combination of 8 dipoles in a simple cubic arrangement leads to a hitherto unheard-of multipole -- a 32-pole or dotriacontapole.

The exploration of the cuboidal dipole arrangement discussed here is triggered by the investigation of magnetic nanoparticles, which have been reported to self-assemble into such configurations \cite{Mehdi2015,Rosenfeldt2018}. The most elementary cluster of this type contains only 8 particles. It can also be assembled macroscopically as a cubic cluster from 8 magnetic spheres, as indicated by the left hand side inset of Fig.\,\ref{fig:B(r)_lin}, and described in Refs.\,\cite{Schoenke2015prb,Borgers2018}. The ground state of this arrangement is stable, and an interesting continuum \cite{Belobrov1983,Schoenke2015prb}. In this state, the spheres attract each other by the magnetic interaction, and in that sense the arrangement can be considered almost self-assembled. 

\section{Experimental results}
\begin{figure}[ht]
\centering
\includegraphics[width=\linewidth]{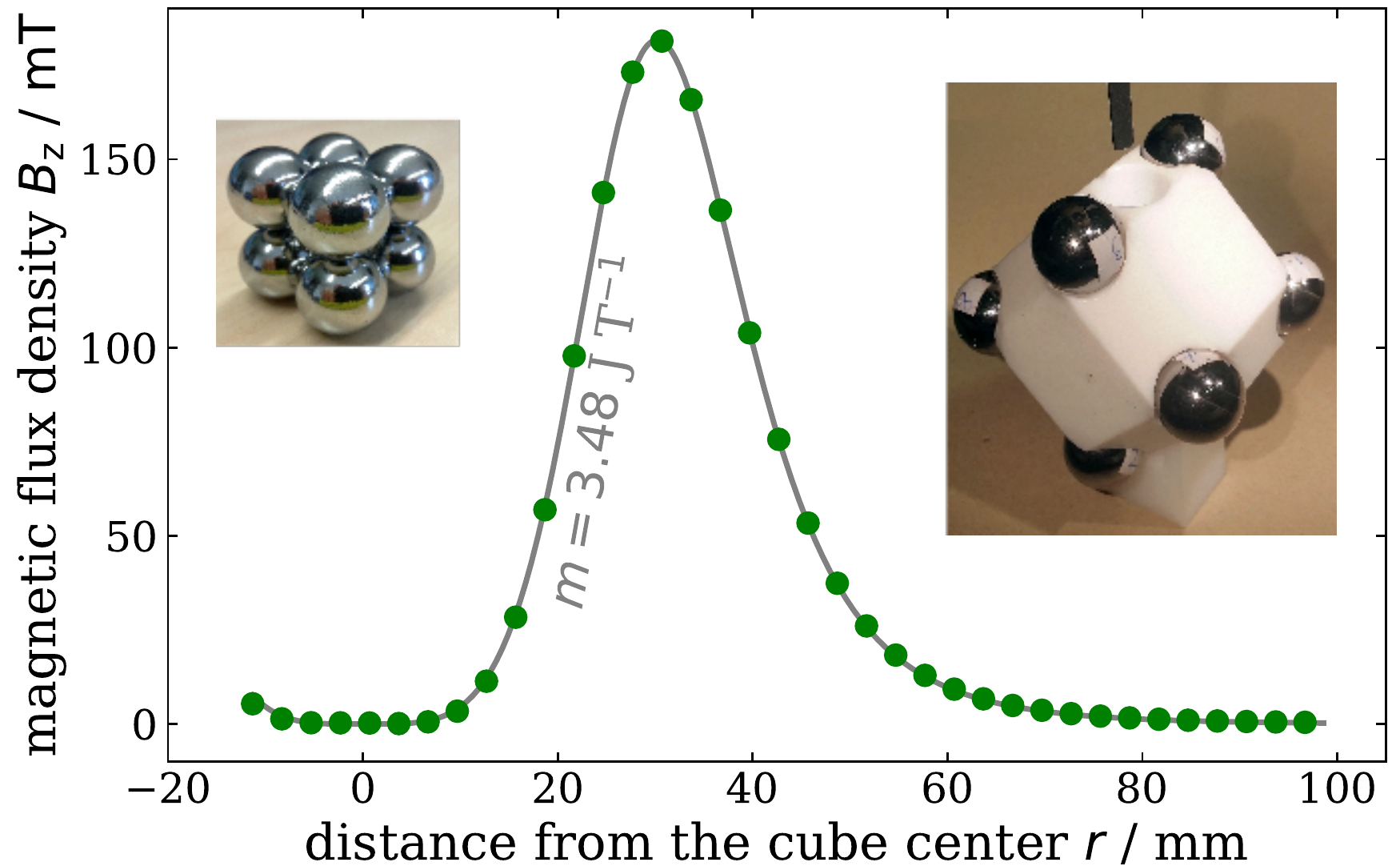}
\caption{Magnetic flux density as measured along a straight path through the center of the cuboid. Only every 30th data point is shown. The solid line corresponds to the numerical superposition of the flux densities of 8 accordingly arranged point dipoles with a magnetic moment $m=3.48\,\mathrm{J\,T^{-1}}$. The left hand side inset shows the principal cuboidal arrangement of the 8 magnetic spheres, and the right hand side inset a geometrically similar arrangement, but here with a white Teflon\textsuperscript{\textregistered} spacer. The hole in that spacer allows to take data inside the cuboid by means of the Hall probe, which is visible as the black part above the hole.}
\label{fig:B(r)_lin}
\end{figure}

For reaching the ground state of the cluster, the eight spheres should be allowed to rotate freely. For that purpose it is useful to provide a Teflon\textsuperscript{\textregistered} spacer to reduce the friction of the spheres, as shown in the right hand side inset of Fig.\,\ref{fig:B(r)_lin}. Here, the eight neodymium magnets of diameter $d=(19\pm 0.05)\,\mathrm{mm}$ are arranged in a cuboidal configuration by the holes at the corners of the white Teflon\textsuperscript{\textregistered} cube, and kept at an edge length $L = (39.5\pm 0.05)\,\mathrm{mm}$ by means of the non-magnetic Teflon\textsuperscript{\textregistered} spacer. A hole is drilled into that spacer along the face diagonal, the (1,1,0) direction of the cube. This allows to move the Hall probe (the black tip) into the cuboid, down to its center, by means of a stepper motor, using 0.1\,mm steps. We adjust the spheres within their continuous ground state to maximize the measured magnetic flux density. This is achieved by manually turning just one sphere around the space diagonal as rotation axis, the other ones follow accordingly due to the magnetic interaction.

The measured magnetic flux density along the (1,1,0) direction is shown in Fig.\,\ref{fig:B(r)_lin}. It has a maximum at about $r=28\,\mathrm{mm}$ -- where the Hall probe is closest to the spheres -- and decays to zero both when approaching the center, and when increasing the distance from the cube. The solid line corresponds to a fit of the numerical superposition of the flux densities of 8 accordingly arranged point dipoles, as given by \eqref{potential} discussed below.

\begin{figure}[ht]
\centering
\includegraphics[width=\linewidth]{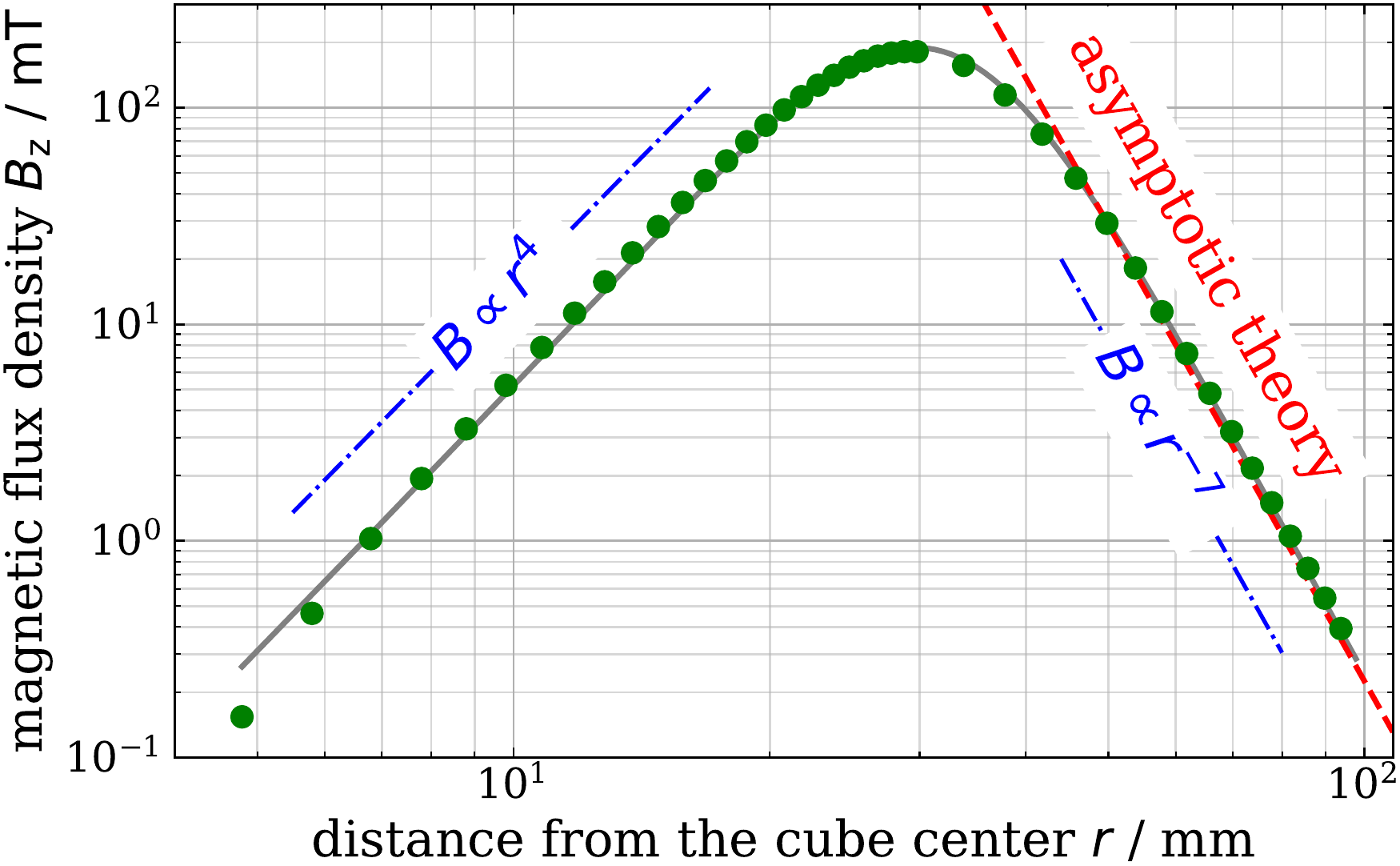}
\caption{The magnetic flux density measured along a path starting from the center of the cuboid is represented  by the circles. Only every 10th data point is shown at the left hand side of the maximum, and every 40th data point at the right hand side. The solid line is the same numerically obtained curve as in Fig.\,\ref{fig:B(r)_lin}. The dash-dotted lines are for comparison with the expected asymptotic slopes. The dashed line depicts the analytical solution \eqref{B(r,tau)} for the far field.}
\label{fig:B(r)_log}
\end{figure}

The most important feature of this cuboidal arrangement of dipoles is the unusually steep decrease of the magnetic flux density outside the cube. To quantify this decrease, Fig.\,\ref{fig:B(r)_log} shows the data from Fig.\,\ref{fig:B(r)_lin} in a logarithmic plot. 
It becomes obvious that the magnetic flux density decays with the inverse 7th power of the distance. To characterize this magnetic cluster with an appropriate name, it must be recalled that the field of dipoles decays with the 3rd power, quadrupoles with the 4th power, and so on. In that sense, the 7th power corresponds to a dotriacontapol. The increase of the flux density with the 4th power near the center is less exotic, however, and reminiscent of the field in a Helmholtz pair of coils. 

\section{Theory}
\begin{figure}[ht]
\centering
\includegraphics[width=.8\linewidth]{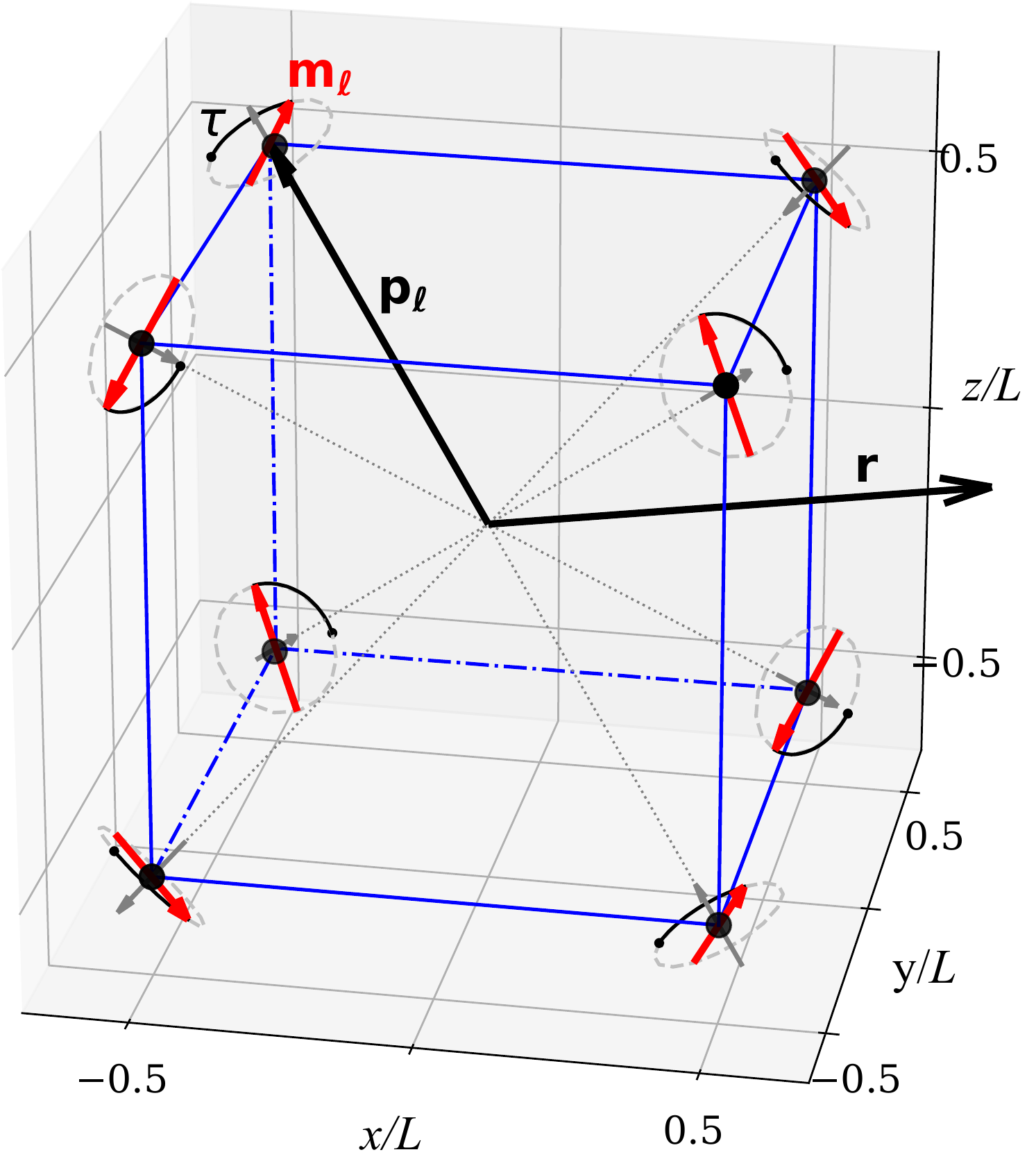}
\caption{The position $\mathbf{r}$ and the position vectors $\mathbf{p}_\ell$ of the dipole moments $\mathbf{m}_\ell$ are taken from the center of the cluster. The orientations of the dipoles in the continuous ground state are determined by the angle $\tau$. The dipole configuration is sketched here for $\tau = 90^\circ$, which corresponds to the largest negative value of $B_\mathrm{z}$ along the (1,1,0) direction.}
\label{fig:3D_sketch}
\end{figure}

To explain the behavior of the magnetic flux density $\mathbf{B}$ in the far field, we perform a \textit{multidipole} expansion. The scalar potential $\phi$ at position $\mathbf{r}$ for a distribution of $N$ dipoles with position vectors $\mathbf{p}_\ell$ and dipole moments $\mathbf{m}_\ell$ (see Fig.~\ref{fig:3D_sketch}) is given by
\begin{equation}
\phi=\sum_{\ell=1}^N\frac{\mathbf{m}_\ell\cdot(\mathbf{r}-\mathbf{p}_\ell)}{4\pi|\mathbf{r}-\mathbf{p}_\ell|^3}
\label{potential}
\end{equation}

The potential is expanded in a series for $|\mathbf{p}_\ell|\ll|\mathbf{r}|$
\begin{equation*}
\phi\;=\;
\sum_{\nu=0}^\infty\frac{1}{\nu!}\sum_{\ell=1}^N
\;\frac{\partial^\nu\phi}{\partial \mathbf{p}_\ell^\nu}\bigg|_{\mathbf{p}_\ell=0}\;\cdot
(\underbrace{\mathbf{p}_\ell\otimes\dots\otimes\mathbf{p}_\ell}_{\nu\mbox{ \footnotesize times}})\,.
\end{equation*}
As an example, the quadrupole (second term in the expansion) reads
\begin{equation*}
\begin{split}
\phi^{(2)}\;&=\;\frac{1}{4\pi|\mathbf{r}|^5}\sum_{\ell=1}^N\Big[3(\mathbf{m}_\ell\cdot\mathbf{r})\mathbf{r}-|\mathbf{r}|^2\mathbf{m}_\ell\Big]\cdot\mathbf{p}_\ell\\
\;&=\;
\frac{1}{4\pi|\mathbf{r}|^5}\sum_{\ell=1}^N
\Big[\underbrace{3\,\mathbf{m}_\ell\otimes\mathbf{p}_\ell-(\mathbf{m}_\ell\cdot\mathbf{p}_\ell)\mathbf{I}}_{\boldsymbol{\mathcal{M}}^2}\Big]\cdot
(\mathbf{r}\otimes\mathbf{r})
\,,
\end{split}
\end{equation*}
with the second order unit tensor $\mathbf{I}$. The second order tensor $\boldsymbol{\mathcal{M}}^2$ is the quadrupole moment. Using Cartesian coordinates $\mathbf{p}_\ell=(p_\ell^x,p_\ell^y,p_\ell^z)$, $\mathbf{m}_\ell=(m_\ell^x,m_\ell^y,m_\ell^z)$, and $\mathbf{r}=(x,y,z)$, we obtain
\begin{equation*}
\begin{split}
\phi^{(2)}\;=\;\frac{1}{4\pi|\mathbf{r}|^5}
\Bigg[&\underbrace{\sum_{\ell=1}^N(2p_\ell^x m_\ell^x-p_\ell^y m_\ell^y-p_\ell^z m_\ell^z)}_{\mathcal{M}^2_{200}}x^2\,+\\
&
\underbrace{3\sum_{\ell=1}^N(p_\ell^x m_\ell^y+p_\ell^y m_\ell^x)}_{\mathcal{M}^2_{110}}xy\,+\,\dots\Bigg]\,.
\end{split}
\end{equation*}
$\mathcal{M}^2_{ijk}$ are the Cartesian components of the moment $\boldsymbol{\mathcal{M}}^2$ with $i+j+k=2$. Using the moments, the potential can be written as
\begin{equation}
\begin{split}
\phi=\sum_{\alpha=1}^\infty\phi^{(\alpha)}=
\sum_{\alpha=1}^\infty\frac{1}{4\pi|\mathbf{r}|^{2\alpha+1}}\hspace{-1pt}\sum_{i+j+k=\alpha}\hspace{-1pt}\mathcal{M}^\alpha_{ijk}\,x^iy^jz^k\,.
\label{potential_expansion}
\end{split}
\end{equation}
The cube ground state \cite{Schoenke2015prb,Schoenke2015gallery} is a highly shielded structure. For a cube with edge length $L$ and dipole moment magnitudes $|\mathbf{m}_\ell|=m$ we have
\begin{eqnarray}
\mbox{(dipole)}         \qquad\mathcal{M}^1_{ijk} &=& 0\nonumber\\
\mbox{(quadrupole)}     \qquad\mathcal{M}^2_{ijk} &=& 0\nonumber\\
\mbox{(octopole)}       \qquad\mathcal{M}^3_{ijk} &=& 0\nonumber\\
\mbox{(hexadecapole)}   \qquad\mathcal{M}^4_{ijk} &=& 0\nonumber\\
\mbox{(dotriacontapole)}\qquad
\mathcal{M}^5_{311} &=& C\,\sin(\tau+\hphantom{5}\pi/3)\label{moments}\\
\mathcal{M}^5_{131} &=& C\,\sin(\tau+5\pi/3)\\
\mathcal{M}^5_{113} &=& C\,\sin(\tau+9\pi/3)\,,
\end{eqnarray}
where $\tau=0\dots 2\pi$ is the current phase angle \cite{Schoenke2015prb,Schoenke2015gallery} of the continuous ground state as indicated in Fig.~\ref{fig:3D_sketch}, and $C=105\sqrt{3/2}\,L^4m$. There are restrictions for the cube moments following from the symmetries of the ground state \cite{Schoenke2015prb}. The potential $\phi$ has to be zero in the three planes $x=0\,, y=0\,, z=0$, as well as on the four volume diagonals $|x|=|y|=|z|$. Together with \eqref{potential_expansion} this leads to conditions for the nonzero moments $\mathcal{M}^\alpha_{ijk}$:
\begin{equation*}
i,j,k\mbox{ positive, odd}\Rightarrow\alpha\mbox{ odd, and}\quad\sum_{i+j+k=\alpha}\mathcal{M}^\alpha_{ijk}=0\,.
\end{equation*}
This explains why the first nonzero moments appear in the dotriacontapole
\begin{equation}
\phi^{(5)}\;=\;\frac{\mathcal{M}^5_{311}\,x^3yz\;+\;\mathcal{M}^5_{131}\,xy^3z\;+\;\mathcal{M}^5_{113}\,xyz^3}{4\pi|\mathbf{r}|^{11}}\,.
\label{dotria}
\end{equation}
The magnetic flux density is related to the potential through $\mathbf{B} = -\mu_0\partial\phi/\partial\mathbf{r}$. We parameterize the measurement along the direction (1,1,0) with the radius parameter $s$ through $(x,y,z)=(s,s,0)/\sqrt{2}$ and obtain the following expression for the $z$-component of the magnetic flux density from \eqref{moments}--\eqref{dotria}
\begin{equation}
\begin{split}
B_z(s,\tau)&=-\mu_0\frac{\partial\phi}{\partial z}\bigg|_{x=y=s/\sqrt{2},\,z=0}\\
&=-\frac{105\sqrt{3/2}\,\mu_0L^4m\,\sin\tau}{16\pi s^7}+\mathcal{O}\Big(\frac{1}{s^9}\Big)\,.
\end{split}
\label{B(r,tau)}
\end{equation}
The next order decays with $|\mathbf{B}|\propto 1/s^9$ because all moments with even $\alpha$ are zero.
%

\eqref{B(r,tau)} is displayed in Fig.\,\ref{fig:B(r)_log} by the dashed line. For the measurements shown there, the angle $\tau$ was adjusted manually to obtain the largest signal of the Hall probe, which correspond either to $\tau=90^\circ$ or to $\tau=270^\circ$. The solid lines in Figs.\,\ref{fig:B(r)_lin} and \ref{fig:B(r)_log} are obtained numerically from the exact \eqref{potential}, with $\tau=90^\circ$ taken as the phase angle of the continuous ground state.

Note that the shape of the $B(r)$-curve shown in Figs.\,\ref{fig:B(r)_lin} and \ref{fig:B(r)_log} are not universal, they rather depend on the direction of the line along which the flux density is measured. The $1/r^7$-decay, however, is a universal feature for all directions in the far field limit $|\mathbf{p}_\ell|\ll|\mathbf{r}|$. 
\section{Conclusion and outlook}

In summary, we have demonstrated that 8 spherical permanent magnets assemble into a configuration which behaves like a dotriacontapole. This can be explained by a model based on pure dipole-dipole interaction. This model is based on symmetry considerations which are an idealization of the experimental situation. The measurements make it clear that the conclusions drawn from the idealization are robust against (small) distortions, in particular the decay of the magnetic flux density with $1/r^7$ -- a hallmark for a highly shielded structure -- survives.

This finding implies that storing strong magnets in a cubic packing might be the optimal way for suppressing their field in the outer surrounding. Moreover, the extremely steep field decay has remarkable consequences for the clustering dynamics: If two dipole spheres, initially separated by say ten diameters, needed one second to collide due to their attractive force, for dotriacontapoles of comparable strength, this process would take more than one year (see Appendix B). Thus, dipoles which manage to arrange themselves in this configuration are fairly robust against further clustering. This argument is scale invariant. It applies to macroscopic granules in the early stages of planet formation \cite{Blum2006} , but could also shed some light on the self-assembly dynamics of colloidal nano-magnets \cite{Mehdi2015,Rosenfeldt2018} used for medical applications \cite{Buzug2010}.

\begin{figure}[ht]
\centering
\includegraphics[width=.8\linewidth]{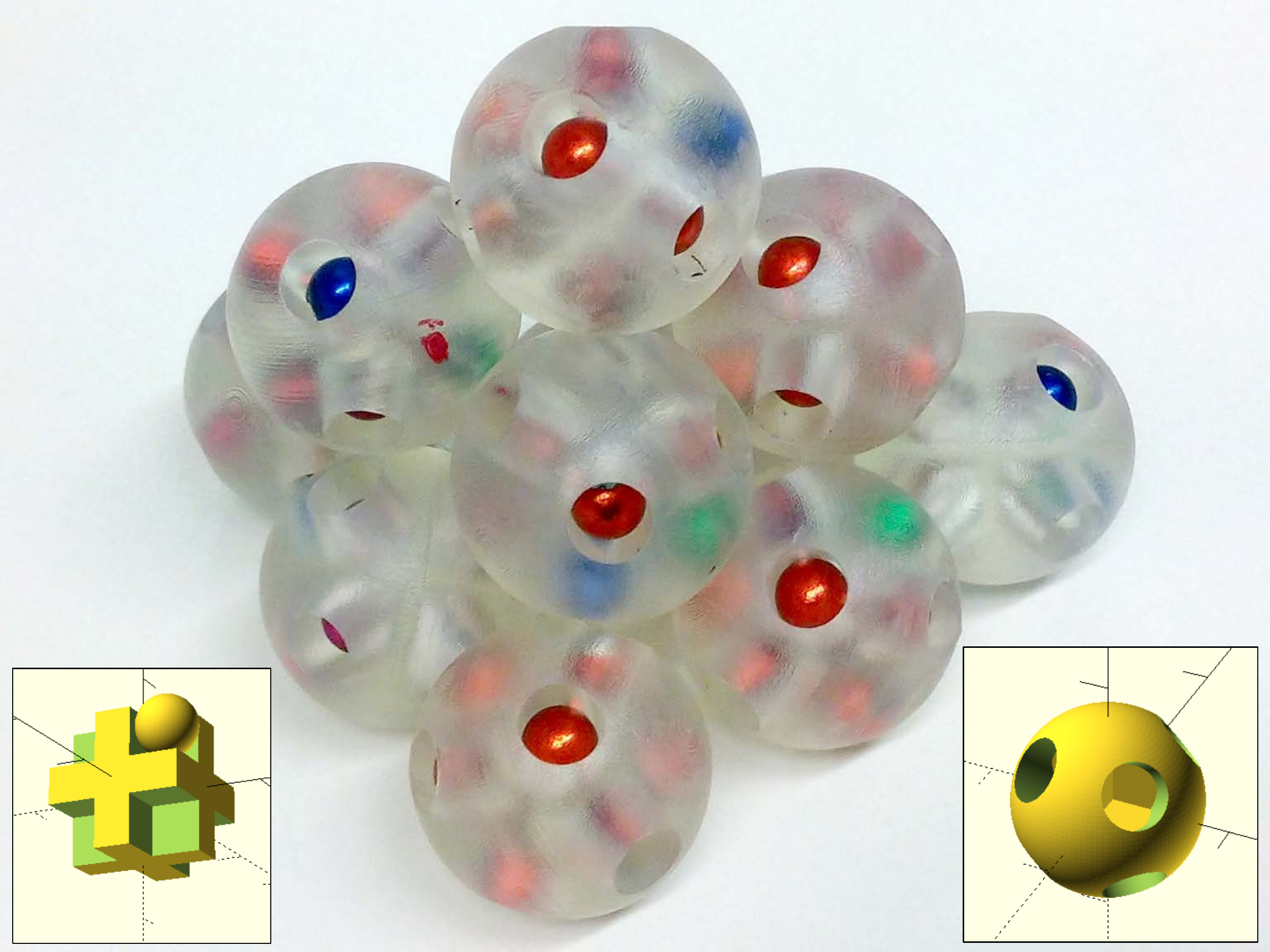}
\caption{A cluster of 3d-printed dotriacontapoles. The inner part of these spheres contains 3 perpendicular walls as indicated by the left hand side inset. The colored magnetic spheres of 5\,mm diameter are placed inside these plastic spheres by the 8 holes along the space diagonals, as indicated by the right hand side inset.}
\label{fig:sphere_cluster}
\end{figure}

The plastic spheres  shown in Fig.\,\ref{fig:sphere_cluster} demonstrate a first attempt to build a handful of such dotriacontapoles with the help of a 3-d printer. Each sphere contains 8 magnetic dipoles in a cubic arrangement. This is provided by 3 perpendicular walls inside these spheres, indicated in the left hand side inset, and 8 holes along the space diagonals, as indicated by the right hand side inset. These plastic spheres should thus interact with an extremely short ranged interaction force, which should asymptotically decay with the inverse 12th power of the mutual distance -- provided that the magnetic dipoles inside a sphere are in their ground state. Measuring such a short range interaction between dipole clusters provides a challenge left to be faced in future work.
\section*{Acknowledgments}
It is a pleasure to thank F.\,Braun, K.\,Huang, R.\,Richter, W.\,Sch\"opf, and A.\,Weber for valuable hints and discussions. This work has been supported by the German Research Foundation (DFG) through grant Re588/20-1.

\appendix
\section{Magnetic spheres as dipoles}
\begin{figure}[b]
\centering
\includegraphics[width=\linewidth]{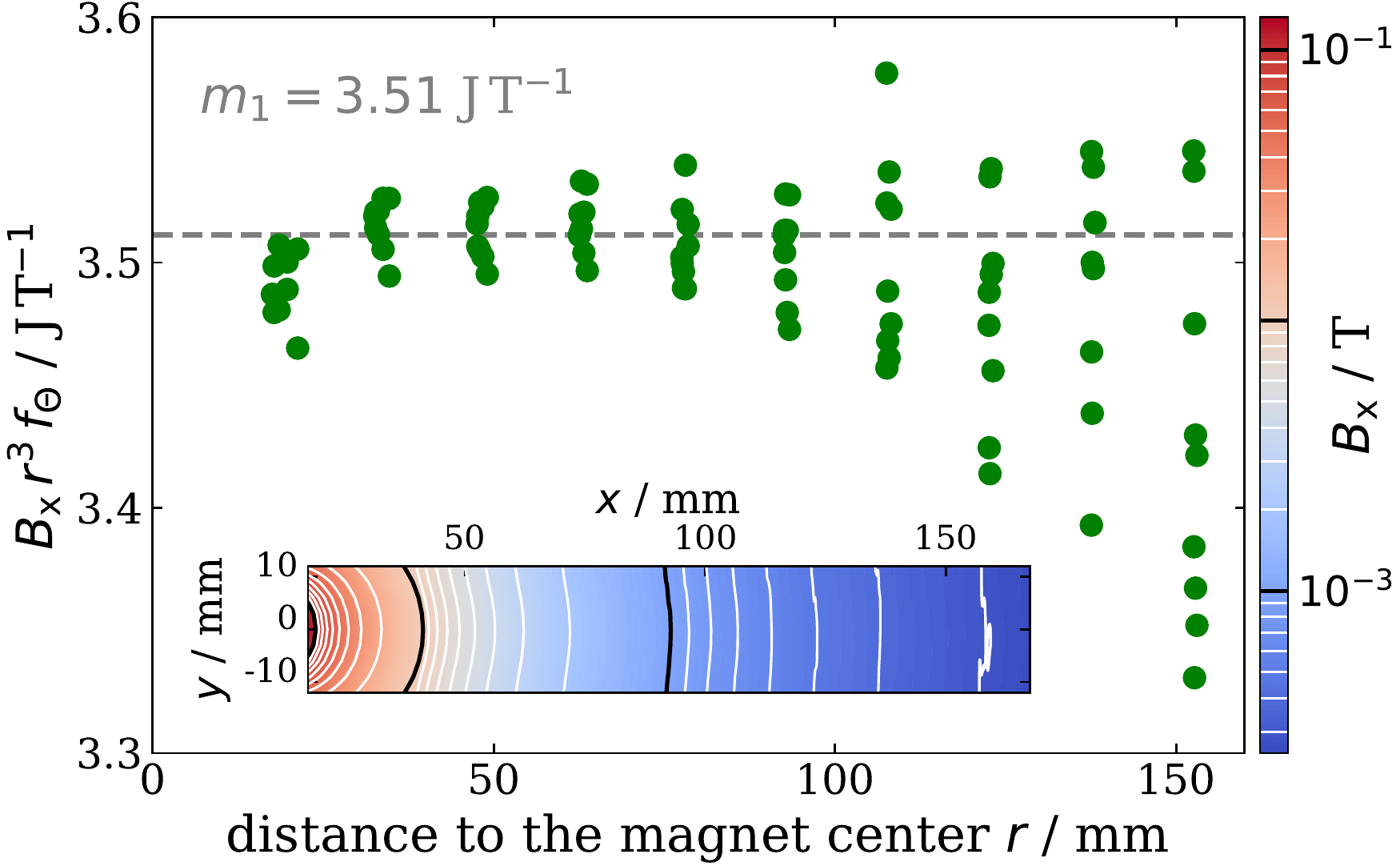}
\caption{The inset shows $B_\text{x}(x,y)$ measured in a plane, color coded in red for strong fields to blue for weak ones. An estimator for the magnetic moment is obtained from these data with \eqref{m(B)}. The result is displayed as a function of $r$ by the circles. The dashed line represents the mean value $m_1$ of these data.}
\label{fig:B(xy)}
\end{figure}

The magnetic spheres (MK-19-C from magnets4you GmbH) have a diameter of $d=(19\pm 0.05)\,$mm. 
For explaining the experimental findings with a theoretical model based on pure dipole-dipole interaction, it is crucial to demonstrate that these spheres can be described as magnetically hard point dipoles. Thus, we have measured the axial component of the magnetic flux density $B_\text{x}$ of a single sphere along the x-direction in a 170\,mm $\times$ 20\,mm xy-plane, as shown in the inset of Fig.\,\ref{fig:B(xy)}. The flux density is measured by a  Hall probe (HU-ST1-184605, MAGNET-PHYSIK Dr.Steingroever GmbH). The 3D-positioning of this probe is done with a stepper motor (High-Z S-400T, with Zero-3 controller from CNC-STEP), the interface (CNCPod) is programmable in G-Code, DIN/ISO 66025. A single-board microcontroller (Leonardo, Arduino) is additionally used for interfacing it to a PC.

To emphasize deviations from the point dipole approximation, and to extract the underlying magnetic moment, we make use of the theoretically expected flux density of a point dipole \cite{Jackson1999}
\begin{equation}
B_\text{x}=\frac{\mu_0}{4\pi}\frac{m\,(3\,\text{cos}^2\Theta-1)}{r^3},
\label{B_x}\end{equation}
with $ \Theta= \arctan(y/x)$, $r=\sqrt{x^2+y^2}$, and the magnetic constant $\mu_0$. With the short hand notation 
$\frac{4\pi}{\mu_0(3\,\text{cos}^2\Theta-1)} = f_\Theta$,
this provides the magnitude of the magnetic moment 
\begin{equation}
m=B_\text{x} r^3 f_\Theta.  
\label{m(B)}\end{equation}
The resulting $m$ as a function of the measured value of $B_\text{x}(x,y)$ is plotted in Fig.\,\ref{fig:B(xy)} as a function of the distance of the Hall probe  from the center of the sphere. The increasing scatter at larger distances $r$ is caused by the fast decay of the magnetic flux density. Based on this data, it seems safe to conclude that the point dipole approximation for the magnetic flux density of the sphere is reliable within $\pm 2\,\%$. The mean value is $(3.51 \pm 0.18)\,\mathrm{J\,T^{-1}}$, which is well within the $(3.54 \pm 0.11)\,\mathrm{J\,T^{-1}}$ claimed by the manufacturer. We have measured all 8 dipoles used in the experiments described here in a similar way, they differ by an amount of $\pm 3\%$.

\begin{figure}[h]
\centering
\includegraphics[width=\linewidth]{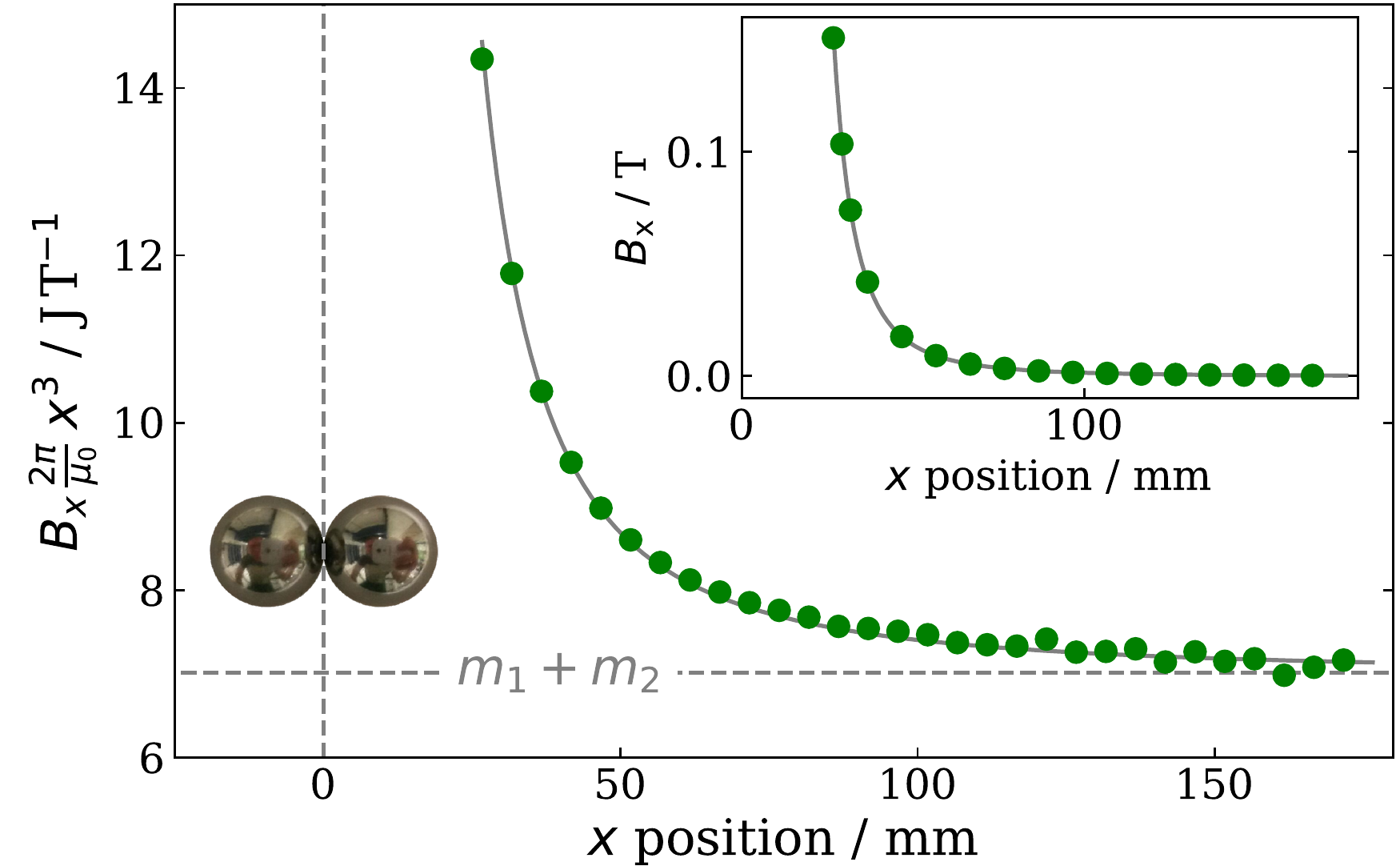}
\caption{The magnetic flux density $B_x$ of two spheres in contact. The raw data are shown in the inset, and the solid line shows the calculated superposition of two dipole fields. $B_x$ scaled with $x^3 2\pi/\mu_0$ is shown in the larger plot. The horizontal dashed line represents the sum of the magnetic moments of the isolated spheres. The dashed vertical line represents the origin at the contact point of the spheres.}
\label{fig:B(r)_two}
\end{figure}

To measure the mutual influence of such magnetic spheres, we brought them in direct contact as shown in the left hand side inset of Fig.\,\ref{fig:B(r)_two}. The measured flux density along the axis of the resulting 2-dipole cluster is shown as the right hand side inset in Fig.\,\ref{fig:B(r)_two}. The position of the Hall probe is measured as the distance from the center between the spheres. 
The data reveal roughly the typical $1/x^3$-descent of a dipole, but deviations from that scaling are hard to judge from this inset plot. To get a better resolution for the deviations from the overall $1/x^3$ decay, the data were multiplied with $x^3$. After scaling with $2\pi/\mu_0$ one gets an estimate for the magnetic moment, which is displayed on the vertical axis of Fig.\,\ref{fig:B(r)_two}. These scaled data decay monotonically with the position $x$ and reach the value of the sum of the two magnetic moments asymptotically, which is indicated by the dashed line. The solid line is the theoretical estimation, based on the superposition of the fields of the individually measured moments $m_1=3.51\,\mathrm{J\,T^{-1}}$ and $m_2=3.50\,\mathrm{J\,T^{-1}}$, with their mutual distance given by the diameter of the spheres. The good agreement between this curve and the data  indicates that the magnets are hard ones: Their magnetic moment stays constant even under the influence of the immediately adjacent other magnet, at least within the experimental resolution on a percentage level.

\section{Assembly time for dipoles versus that for dotriacontapoles}
The time $T_m$ for two multipoles of diameter $d$ starting at a distance of $10\,d$ to come into contact under the influence of their mutual attraction -- a characteristic time for the dynamics of the self-assembly of magnetic clusters \cite{Mehdi2015,Rosenfeldt2018} -- is obtained by integrating over their inverse velocity. When assuming that these particles are suspended in a viscous fluid, that velocity is proportional to the attractive force (Stokes's law).  $T_2$ denotes the pair of dipoles, $T_{32}$ the pair of dotriacontapoles. The attracting force of these multipole pairs is assumed to be the same when they are in contact at the distance of $1\,d$.
\begin{equation*}
\begin{split}\frac{T_{32}}{T_2}=\frac{\int\limits_{\text{5\,d}}^{\text{d/2}} \frac{1}{v_{32}}\,\text{d}r}{\int\limits_{\text{5\,d}}^{\text{d/2}} \frac{1}{v_{2}}\,\text{d}r}&\overset{v\propto F}{=}\frac{\int\limits_{\text{5\,d}}^{\text{d/2}} -\left(\frac{2\,r}{\text{d}}\right)^{12}\,\text{d}r}{\int\limits_{\text{5\,d}}^{\text{d/2}} -\left(\frac{2\,r}{\text{d}}\right)^{4}\,\text{d}r}=\\
&=\frac{5}{13}\,\frac{10^{13}-1}{10^5-1}\approx 0.4\cdot 10^8
\end{split}
\end{equation*}
This ratio turns, e.\,g., 1\,s for a dipole pair into 1\,a for the corresponding pair of dotriacontapoles: They are fairly robust against further clustering. 



\bibliography{Hartung18}

\end{document}